# Orbital Signatures of Density Wave Transition in $La_3Ni_2O_{7-\delta}$ and $La_2PrNi_2O_{7-\delta}$ RP-Nickelates Probed via *in-situ* X-ray Absorption Near-edge Spectroscopy

Mingtao Li[1,*,#], Mingxin Zhang[2,*], Yiming Wang[1], Jiayi Guan[1], Nana Li[1], Cuiying Pei[2], N-Diaye Adama[3], Qingyu Kong[3], Yanpeng Qi[2,4,5,#], and Wenge Yang[1,#]

[1]Center for High Pressure Science and Technology Advanced Research, Shanghai 201203, China

[2]School of Physical Science and Technology, ShanghaiTech University, Shanghai 201210, China

[3]Synchrotron SOLEIL, L'Orme des Merisiers, Saint-Aubin, BP48, 91192 Gif-sur-Yvette, France

[4]ShanghaiTech Laboratory for Topological Physics, ShanghaiTech University, Shanghai 201210, China

[5]Shanghai Key Laboratory of High-resolution Electron Microscopy, ShanghaiTech University, Shanghai 201210, China

[*] These authors contributed equally to this work.

**#Corresponding authors:**

mingtao.li@hpstar.ac.cn (M.T.L.), qiyp@shanghaitech.edu.cn (Y.P.Q.), yangwg@hpstar.ac.cn (W.G.Y.)

## ABSTRACT

The report of superconductivity (SC) with $T_c$~80 K in bilayer Ruddlesden-Popper (RP) nickelate $La_3Ni_2O_{7-\delta}$ have sparked considerable investigations on its normal state properties and SC mechanism under pressure and at low temperature. It is believed that the density wave (DW) at ~150 K plays an important role in SC emergence, but its nature remains largely underexplored. Here, we utilized temperature-dependent *in-situ* Ni K-edge X-ray Absorption Near-edge Spectroscopy (XANES) to probe the Ni-$3d/4p$ electronic states of $La_3Ni_2O_{7-\delta}$ and $La_2PrNi_2O_{7-\delta}$ samples down to 4.8 K, enabling us to witness the evolution of both in-plane $d_{x^2-y^2}/p_x(p_y)$ and out-of-plane $d_{3z^2-r^2}/p_z$ orbitals of $NiO_6$ octahedron across the DW transition. Main edge energy associated with Ni $4p$ orbital shows an anomalous decline near DW transition, signifying the occurrence of lattice distortions as a hallmark of charge density wave. Below DW transition, the enlarged crystal field splitting (CFS) indicates an enhanced $NiO_6$ octahedral distortion. Intriguingly, magnetic Pr substituents could activate the mutual interplay of $d_{x^2-y^2}$ and $d_{3z^2-r^2}$ orbitals. We discussed its relevance to the favored bulk SC in the pressurized polycrystalline $La_2PrNi_2O_{7-\delta}$ than pristine.

**Subject Areas:** Condensed Matter Physics, Strongly Correlated Materials, Superconductivity

## I. INTRODUCTION

High-$T_c$ SC favors to appear at the boarder of the complex electronic/magnetic ordered states [1], such as antiferromagnetic (AFM), spin density wave (SDW), and charge density wave (CDW) orders. Understanding of their mutual interplay has been one central theme in condensed matter physics. Recent discovered high-$T_c$ SC at ~80 K in pressurized bilayer RP-nickelates have been attracting considerable interests in elucidating its SC mechanism [2,3]. Similar to the phase diagram of cuprates, SC emerges after suppressing a DW transition by applying pressure in bilayer $La_3Ni_2O_{7-\delta}$ (La327) and trilayer $La_4Ni_3O_{10-\delta}$ (La4310) RP-nickelates [4-6]. La327 features a strong hybridization between $d_{3z^2-r^2}$ and $d_{x^2-y^2}$ orbitals through the intralayer and interlayer Ni-O-Ni bonds [7]. The $d_{3z^2-r^2}$ and $d_{x^2-y^2}$ orbitals pronouncedly hybridize with oxygen $2p$ orbitals as well. This distinguishes from the high-$T_c$ cuprates with dominant in-plane $d_{x^2-y^2}$ and oxygen $2p$ orbital hybridization [8,9], forming the Zhang-Rice singlet [10].

DW is the periodical modulation either from electron's charge or spin widely observed in conventional metals and strongly correlated materials. DW transition of La327 has been investigated by various experimental probes. Early magneto-transport measurements show that DW anomaly occurs at ~120 K with different oxygen stoichiometry from temperature-dependent resistivity $\rho(T)$ and Hall coefficient $R_H(T)$ [11,12]. Recent $\rho(T)$ results on a La327 single crystal reveal two anomalies at about 110 K and 153 K [13], indicating the formation of CDW and SDW orders. Spectroscopic measurements of muon spin ration spectroscopy (μSR) [14], resonant inelastic X-ray spectroscopy (RIXS) [15], resonant X-ray scattering (RXS) [16,17], ultrafast optical spectroscopy [18], and nuclear magnetic resonance (NMR) spectroscopy [19-21] have clearly revealed a SDW transition at ~150 K in ambient La327 bulk/thin film single crystals. Ultrafast optical spectroscopy evidences two DW transitions in pressurized La327 single crystal [18], which is suggested to be SDW type for $P$ <26 GPa and CDW type for $P$ >29.4 GPa. Theoretical calculations demonstrate that La327 hosts an AFM ground state [22], and strong Fermi surface nesting can lead to phonon softening and electronic instabilities to form CDW orders at low and high pressures. Recent μSR study uncovers two distinct DW transitions with a pressure-induced separation of SDW order at higher temperature and a proposed CDW order at lower temperature [23], further casting the mystery nature of DW transition in La327.

In La4310, an intertwined state forms below DW transition temperature ($T_{\mathrm{DW}}$) at ~140 K [24], possessing both charge and spin order. Currently, it is still unknown whether CDW concomitantly exists in La327 or not. In addition, bulk SC is reported in pressurized $La_2PrNi_2O_{7-\delta}$ polycrystalline (denoted as La2Pr327) [25], while absent in pristine [4]. A comparative study of the electronic structure associated with Ni $3d$ orbitals is lack yet, which is important for understanding its emerged bulk SC mechanism through Pr-substitution. This motivates the present study of orbital signatures across DW transition in La327 and La2Pr327 using temperature-dependent XANES at Ni $K$-edge, which is a spectroscopy probe with both electronic and lattice sensitivity from pre-edge and main edge absorption. It thus enables one to investigate the orbital signatures of Ni-$3d/4p$ and their mutual interplay in RP-nickelates. Polycrystalline samples of La327 and La2Pr327 are synthesized by sol-gel method [4], and XANES data were collected at ODE beamline [26], synchrotron SOLEIL. More details on synthesis, XANES measurements and theoretical simulations can be found in ref. [27].

## II. RESULTS

### A. Crystal structure and DW anomaly

To compare lattice parameters and distortion parameter of $NiO_6$ octahedron, we conducted powder XRD measurements with Rietveld structural refinements of both La327 and La2Pr327 nickelates to retrieve the crystal structure by GSAS II package [28]. Refined profiles are displayed in Figs. 1(a) and (b). Obtained parameters are summarized in table S1 [27]. Both samples are phase pure and have orthorhombic structure with $Amam$ symmetry, consistent with previous reports [25,29]. After Pr-substitution, volume shrinks due to smaller ion radius of $Pr^{3+}$ (1.126 Å) than $La^{3+}$ (1.16 Å) with coordinating eight oxygen atoms [30]. Meanwhile, Ni-O-Ni bond angle ($\alpha$) and interlayer Ni-O-Ni bond angle ($\beta$) of $NiO_6$ octahedral tend to decline in La2Pr327, indicating weakened interlayer coupling of $d_{3z^2-r^2}$ orbital through hybridizing with oxygen's $p_z$ orbital. To unveil Pr-substitution effect on $NiO_6$ octahedral distortion, we calculated distortion parameter $\Delta$ of a $MO_N$ polyhedron ($M$ is the central atom coordinated by $N$ oxygens) [31], defined as $\Delta = (1/N)\sum_{n=1,N}\{(d_n - \langle d \rangle)/\langle d \rangle\}^2$ with $n$ is an integer denoting $n^{\text{th}}$ M-O distance and $N$ coordination number. Unexpectedly, $\Delta$ decreases after Pr-substitution (0.00352/0.00322 for La327/La2Pr327, see table S1). This implies a reduction of $d_{x^2-y^2}$ orbital and $d_{3z^2-r^2}$ orbital energy splitting, as validated by the following XANES results.

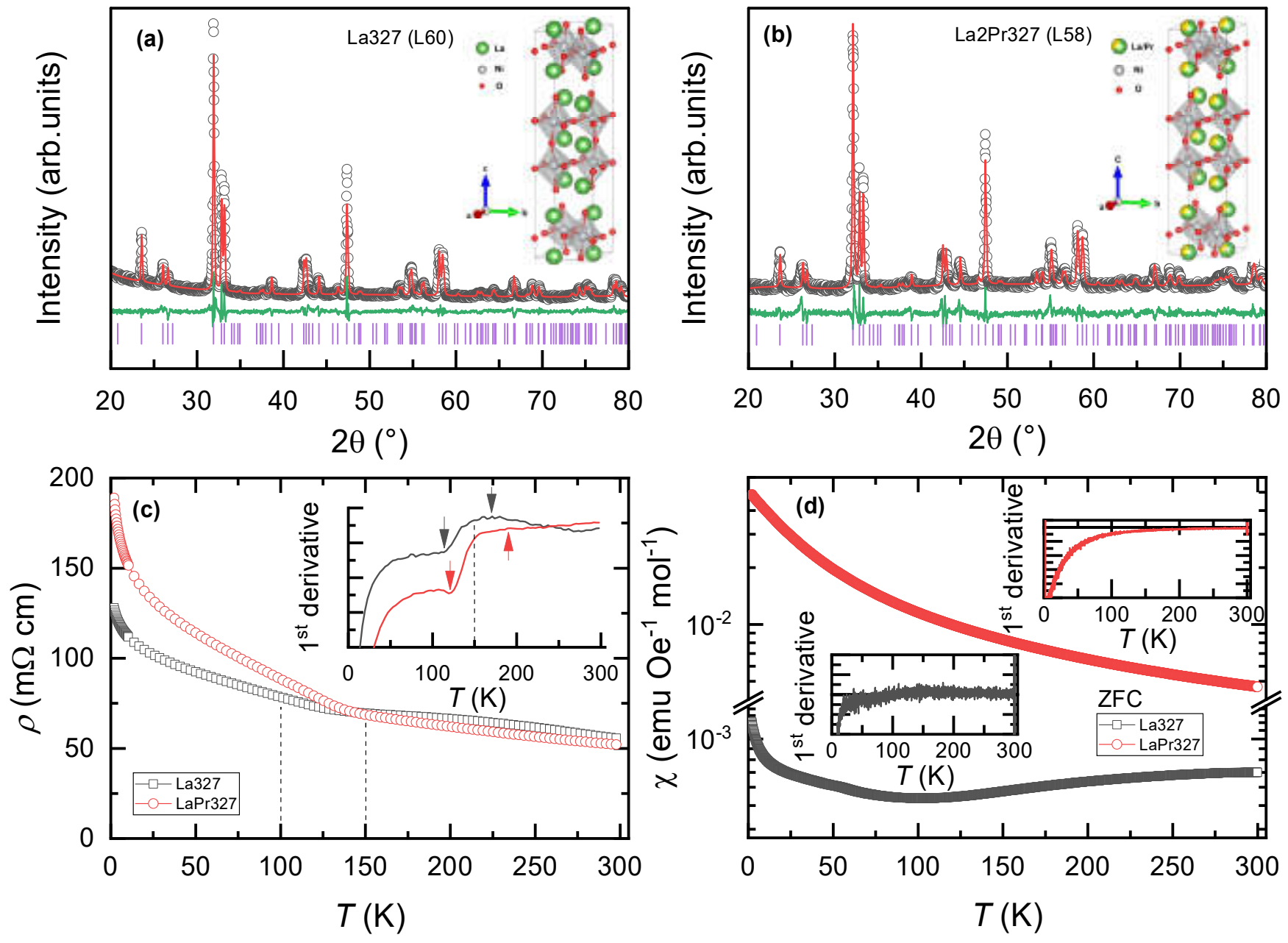

**FIG. 1. Crystal symmetries and DW transitions in bilayer La327 and La2Pr327 RP-nickelates.** Powder XRD patterns and Rietveld structural refinements of (a) La327 and (b) La2Pr327 at ambient conditions. (c) $\rho(T)$ curves. Inset shows its first derivative with DW anomalies indicated by arrows. (d) Temperature-dependent magnetization susceptibility $\chi(T)$. Inset shows its first derivative.

DW anomalies were reported previously in La327 [4,11,13,32,33], which manifests kinks in $\rho(T)$ and $\chi(T)$ curves between 100-150 K. As seen in Fig. 1(c), there occurs a clear resistivity anomaly below ~150 K for both samples. First derivatives show that two extrema occur at ~171(113) K for La327 and ~190(120) K for La2Pr327. Spin fluctuations above 150 K have been experimentally observed in $La_3Ni_2O_{7-\delta}$ by μSR [14], NMR [21] and RIXS [15]. We can attribute the resistivity anomalies above 150 K to spin fluctuation induced scattering rates variation of itinerant carriers. Interestingly, $\chi(T)$ shows a contrasting behavior. For La327, $\chi(T)$ severely deviates from Curie-Weiss (CW) law with a dip at ~100 K, in agreement with other reports [11,13,14,32]. Nevertheless, $\chi(T)$ of La2Pr327 approximately obeys CW behavior. A simple fitting of modified CW law [34], $\chi(T) = \chi_0 + C/(T - \Theta_{\rm w})$, yields a temperature-independent term $\chi_0$=4.5(2)×$10^{-4}$ $cm^3$/mol, Curie constant $C$=1.596(4) $cm^3$ K/mol, Weiss temperature $\Theta_{\rm w}$=-29.41(8) K. Fitting profile can be found in Fig. S1 [27]. Negative value of $\Theta_{\rm w}$ indicates a net AFM exchange interaction between formal $Ni^{2.5+}$ spins and/or $Pr^{3+}$ spins. An effective spin moment can be further estimated by $\mu_{\rm eff} \approx \sqrt{8C}$, yielding 3.57 $\mu_B$. This is significantly larger than La327 with moment $\cong 0.3 - 0.7$ $\mu_B$ per Ni [23], but very close to an isolated $Pr^{3+}$ moment 3.58 $\mu_B$ with $^3F_4$ configuration. Accordingly, the paramagnetic behavior of La2Pr327 is most likely contributed by magnetic $Pr^{3+}$ ions instead of Ni ions. In addition, no apparent anomalies are observable in their first derivative besides a weak kink at ~150 K for La327 [insets of Fig. 1(d)].

**B. XANES: experimental data and theoretical simulations**

Through Ni $K$-edge XANES measurements [35], one can not only obtain the orbital signatures of unoccupied Ni-3$d$ electronic states near Fermi energy $E_{\rm F}$ from analyzing the pre-edge peak but also detect the subtle lattice distortion evidenced from the main edge for the unoccupied Ni-4$p$ electronic states. As shown in Fig. 2(a), there are five features, $a$, $A$, $B'/B$ and $C$, observable on XANES spectra for polycrystalline La327 and La2Pr327. Specially, it is demonstrated that the pre-edge $a$ peak arises from a minor quadruple contribution (<5%) of $1s \rightarrow 3d$ transitions, and dominant dipole allowed $1s \rightarrow 4p$ transitions due to enhanced $p$ orbital component of strong Ni-O orbital hybridizations. While, doublet main edges $B'/B$ are interpreted by the CFS due to ligand oxygen atoms [see Fig. 2(b)] and core-hole screening of different $3d$ configurations, i.e., $1s \rightarrow 4p_{\pi}3d^8\underline{L}$ / $1s \rightarrow 4p_{\sigma}3d^8\underline{L}$ for out-of-plane and in-plane contributions. $A$ and $C$ peaks are attributed to $1s \rightarrow 4p_{\mathrm{hyb}}^{A}$ transition for the enhanced $4p$ component due to $p-d$ hybridization of La and O atoms, and transition of $1s \rightarrow 4p3d^8$, respectively. To compare mutual difference, we have conducted theoretical simulations of Ni $K$-edge XANES spectra with considering spin-orbital coupling (SOC) using FDMNES code [36,37], which is mainly a fully relativistic DFT-LSDA code allowing *ab initio* simulations [38]. Together with experimental XANES spectra, corresponding first and second derivatives, we plot them in Figs. 2(a), Figs. 2(a') and (a'') for comparisons.

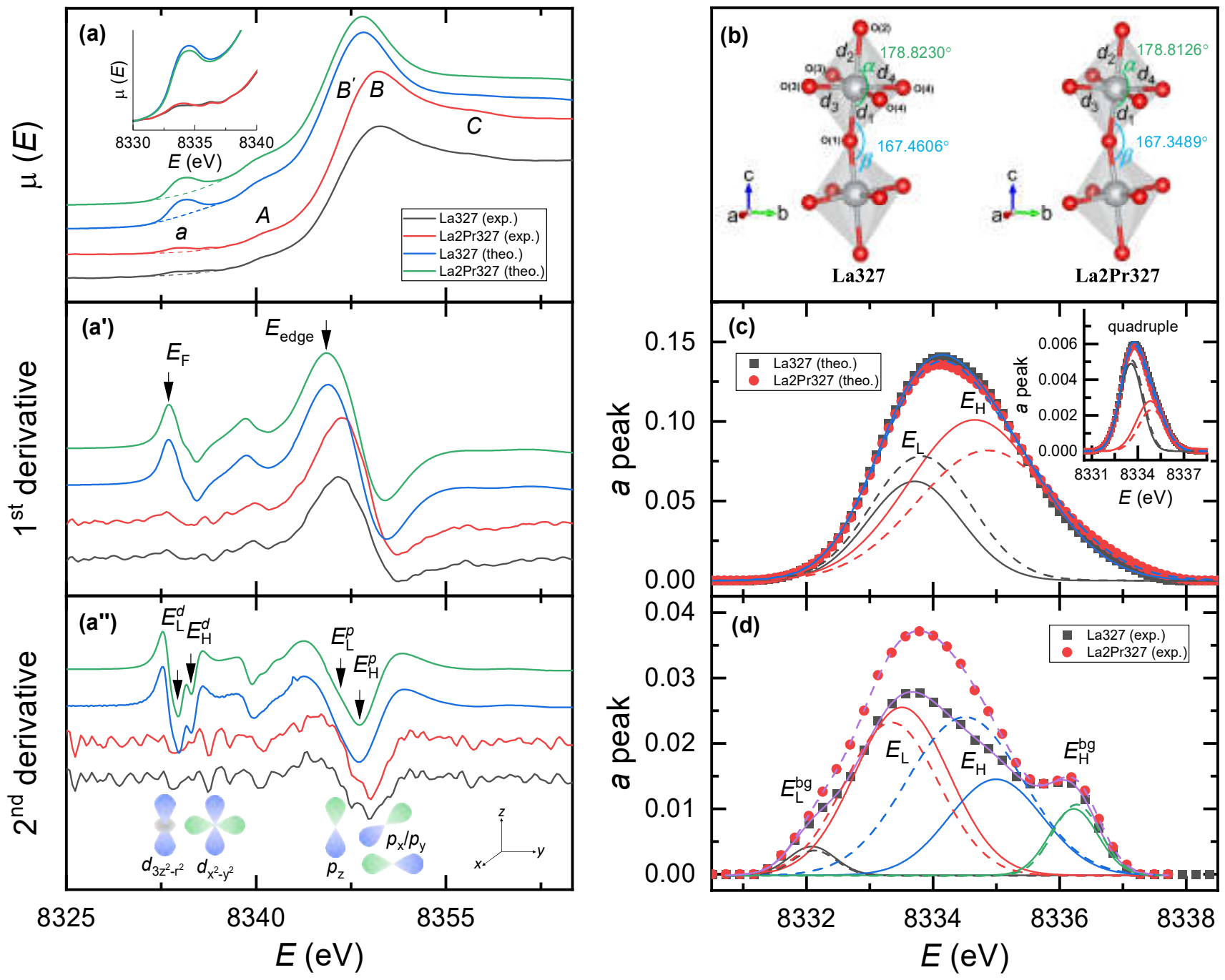

**FIG. 2. Theoretical and experimental comparisons of Ni $K$-edge XANES spectra.** (a) Experimental and theoretical simulated XANES spectra. Alphabets of $a$, $A$, $B'$, $B$ and $C$ indicate main features. (a')-(a'') are corresponding first-order and second-order derivatives. Arrows indicate edge and peak positions. Insets of (a'') show corresponding orbitals. (b) Schematics of bilayer $NiO_6$ octahedra with Ni-O bond distance $d_3 < d_4 < d_1 < d_2$. (c) and (d) are theoretical and experimental $a$ peak fitted by Gaussian functions. Inset of Fig. (c) shows quadruple contribution. Experimental data are collected at 300 K.

To determine $E_{\mathrm{F}}$, main edge $E_{\mathrm{edge}}$, $E_{\mathrm{L}}$ and $E_{\mathrm{H}}$ for lower and higher peak, and CFS energy $\Delta E = E_{\mathrm{L}} - E_{\mathrm{H}}$, we have fitted corresponding first-order and second-order derivatives of XANES spectra approximated by Voigt function [see arrows in Figs. 2(a') and (a'')]. Full XANES spectra can be found in Fig. 5. Simulated and experimental $\Delta E^{d}$ values for pre-edge $a$ peak was obtained by Gaussian function fitting [see Figs. 2(c) and (d)]. Extracted $a$ peak and $A$ peak at various temperature are displayed in Fig. S2 [27]. Pure quadruple contribution due to $1s \rightarrow 3d$ transitions is plotted as inset of Fig. 2(c), and Gaussian fitting gives $\Delta E$ values of 1.26(6)/1.32(5) eV for La327/La2Pr327. Due to constant background, two additional Gaussian functions are included for fitting experimental $a$ peaks. We summarized fitting results in table S2 [27]. It is found that simulated $\Delta E$ values for $d$ orbitals and $p$ orbitals are about 0.5 eV lower and higher than those of experimental. This probably arises from Hubbard $U$ term and ligand-hole effect not considered in our FDMNES calculations. Regardless of this, $\Delta E_{\mathrm{exp.}}$ values are quite close, indicating they are good indicators for characterizing $NiO_6$ octahedral distortion. As pointed out afore [35], $\Delta E_{\mathrm{exp.}}$ mainly corresponds to energy splitting between $d_{3z^2-r^2}$ and $d_{x^2-y^2}$ orbitals, and degenerated $4p_x/4p_y$ and $4p_z$ orbitals splitting. The experimental fact that $\Delta E^{d}$ and $\Delta E^{p}$ values of La2Pr327 are smaller than those of La327 strongly supports that Pr-substitution could averagely reduce $NiO_6$ octahedra distortion.

### C. Main edge energy $E^{p}_{\mathrm{edge}}$ and $p$-orbital energy splitting

To study the nature of DW transition at ~150 K, we have extracted temperature-dependent $E^{p}_{\mathrm{edge}}$, $E^{p}_{\mathrm{L}}$, $E^{p}_{\mathrm{H}}$, and $\Delta E^{p}$ at main edge, which is associated to $1s \rightarrow 4p$ transitions [see Figs. 3(a)-(c)]. Energy splitting of $p$ orbitals in La327 are previously confirmed [35]. According to Natoli's rule [39], $E^{p}_{\mathrm{edge}}$, $E^{p}_{\mathrm{L}}$, and $E^{p}_{\mathrm{H}}$ are sensitive to structural factor. This assures that XANES can serve as a probe of CDW or symmetry breaking induced subtle lattice distortion, which readily causes energy anomalies. As seen in Fig. 3(a), $E^{p}_{\mathrm{edge}}$ of La2Pr327 is indeed higher than that of La327, caused by larger lattice contraction for Pr-substitution. With temperature decreases to 175 K, $E^p_{\text{edge}}$ starts to abnormally deviate from the linear increase trend due to cooling-induced lattice shrinking. Decrease of $E^p_{\text{edge}}$ across DW transition evidences that lattice expansion occurs for both samples. Recent single crystal diffraction study demonstrate that in-plane $b$ and out-of-plane $c$ lattice parameters show anomalous expansion below 200 K and $Amam$ symmetry is enhanced [40]. More details can be found in ref. [40] and in Fig. S3 and Fig. S4 in ref. [27]. Together with absence of symmetry breaking, our XANES results support DW transition is also of CDW nature in La327 and La2Pr327.

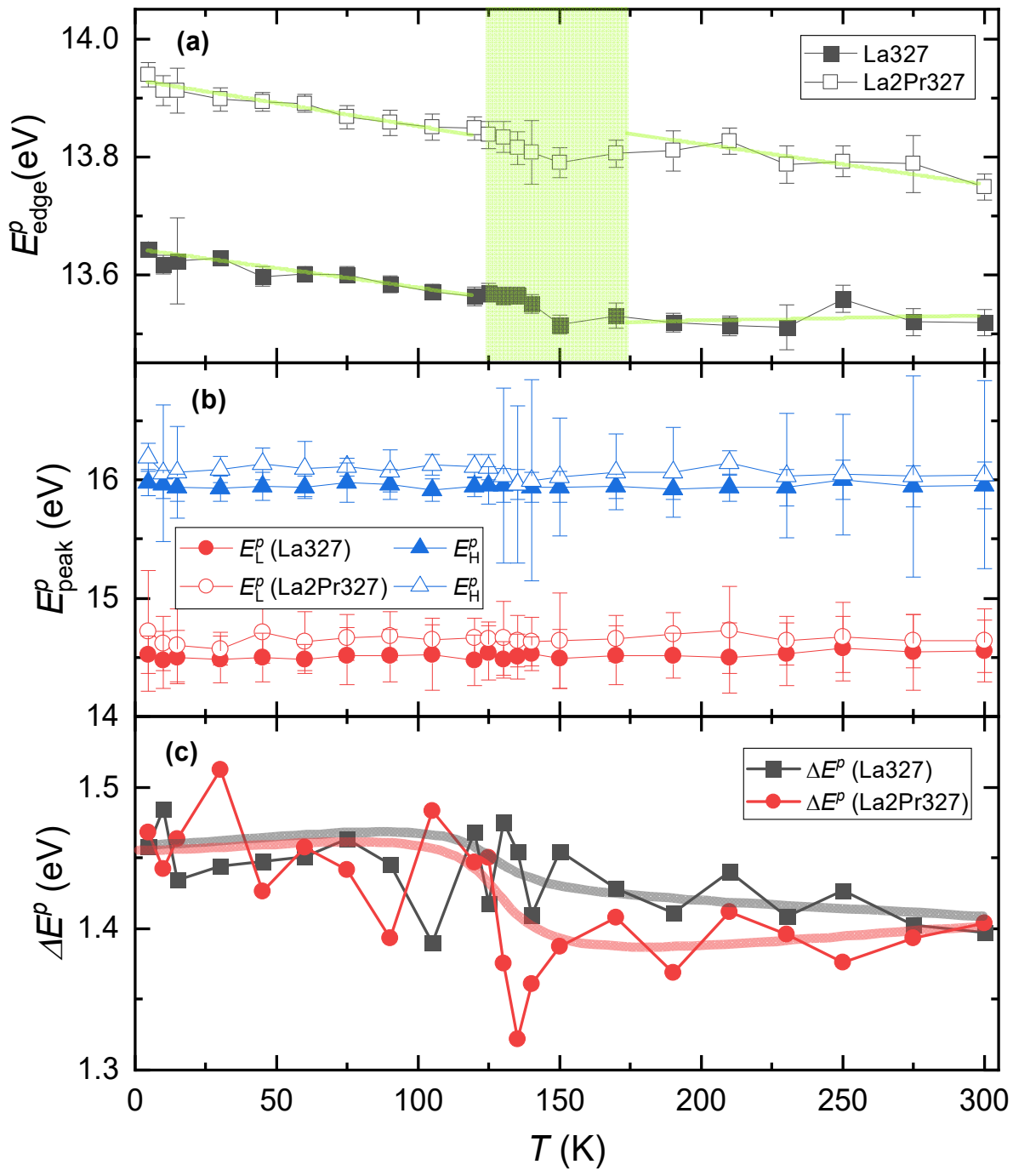

**FIG. 3. Experimental evidences of DW-induced enhancement of $p$-orbital energy splitting.** (a) Main edge energy $E^p_{\text{edge}}$ guided by solid linear fitting lines. (b) Temperature-dependent $E^p_{\text{L}}$, $E^p_{\text{H}}$, and (c) $\Delta E^p$ of $B'/B$ peak for La327 and La2Pr327. All energy values are referred to Ni foil edge $E_0$ =8333 eV. Solid symbols are for La327, and empty symbols for La2Pr327. Bold lines in Fig. (c) are guides to the eye.

To investigate how $NiO_6$ distortion responses to DW transition, values of $E^p_{\text{L}}$, $E^p_{\text{H}}$, and $\Delta E^p$ are extracted, which approximately correspond to energy position for out-of-plane $p_z$ orbital ($\pi$-polarization, $\vec{e} \parallel c$) and in-plane $p_x/p_y$ orbital ($\sigma$-polarization, $\vec{e} \perp c$), and CFS energy. Slightly higher $E_{\mathrm{L}}^{p}$ of La2Pr327 can be well understood from its smaller $c$-axis lattice parameter while mean in-plane lattice parameters are comparable [see Fig. 3(b)]. Above $T_{\mathrm{DW}}$, lower $\Delta E^{p}$ is well consistent with the smaller $\Delta$ value of $NiO_6$ in La2Pr327 [see Fig. 3(c)]. Below $T_{\mathrm{DW}}$, $\Delta E^{p}$ values tend to increase and become comparable, signifying an enhanced $NiO_6$ distortion. With decreasing temperature, in-plane and out-of-plane bond angles of Ni-O-Ni show much stronger deviation from 180° in ambient La327 single crystal [40], showing enhanced $NiO_6$ octahedral tilts/distortion and increasingly favorable $Amam$ symmetry. From $\Delta E^{p}(T)$ data of La2Pr327, we also note that it undergoes severer thermal response than La327, signaling a promoted thermal orbital sensitivity after Pr-substitution.

**D. Fermi energy shift and $d$-orbital energy splitting**

Since the physics of nickelates are mostly contributed by $3d$ orbitals near $E_{\mathrm{F}}$, we have investigated temperature evolution of $a$ peak that defines $E_{\mathrm{F}}$ as the maximum of its first derivative. Firstly, we plot temperature-dependent $E_{\mathrm{F}}$ in Fig. 4(a). Angle resolved photoemission spectroscopy (ARPES) measurements have revealed one electron-like pocked and one hole-like pocket originated from the hybridized itinerant Ni $d_{x^2-y^2}$ orbital and oxygen's $p_{\mathrm{x}}/p_{\mathrm{y}}$ orbitals [41]. The $d_{3z^2-r^2}$ orbital derived bands locate below $E_{\mathrm{F}}$ for bonding one and above $E_{\mathrm{F}}$ for antibonding one, which are more localized due to much stronger electronic correlation strength [42,43]. To check how $d$-orbital splitting evolutes across DW transition, we summarized fitting results of $E_{\mathrm{F}}$, $E_{\mathrm{L}}^{d}$, $E_{\mathrm{H}}^{d}$, $\Delta E^{d}$ and peak area in Figs. 4(a)-(c). $E_{\mathrm{F}}$ tends to decrease for both samples but with a higher rate for La327 than La2Pr327, an indicative of increasing electron counts of Ni-$3d^{7.5}$. Below ~150 K, a notable increase of $E_{\mathrm{F}}$ occurs for La327 while decrease for La2Pr327, manifesting a contrasting thermal response between 120 K and 150 K. This implies a distinguishing electronic response across DW transition. To further investigate $d$-orbital splitting, we plotted $E_{\mathrm{L}}^{d}$, $E_{\mathrm{H}}^{d}$, and $\Delta E^{d}$ data in Fig. 4(b). Intriguingly, all of them are almost thermal inert for La327, but show observable temperature-dependence for La2Pr327. Especially, $\Delta E^{d}$ undergoes a dip across DW transition for La2Pr327, in consistent with $\Delta E^{p}$ extracted from $p$ orbitals [see Fig.

3(c)]. This verifies that electronic anomaly definitely accompanies with lattice distortion, which varies CFS. Also, smaller $\Delta E$ of La2Pr327 above $T_{DW}$ again confirms a relative weaker $NiO_6$ distortion.

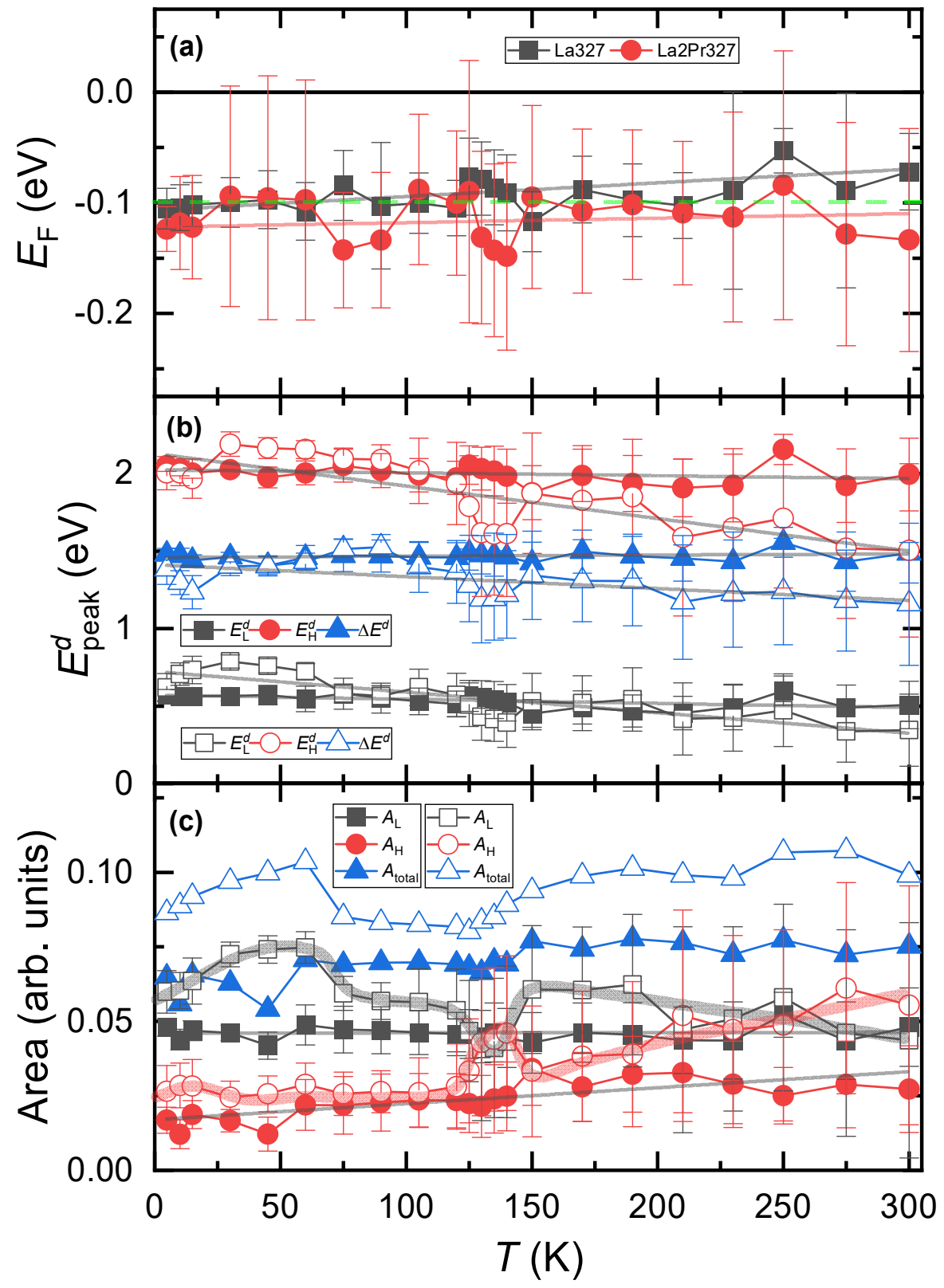

**FIG. 4. Experimental evidences of DW-induced anomalies of $d$-orbital energy splitting.** (a) Temperature-dependent relative $E_F$ for La327 and La2Pr327. (b) Peak energy $E^d_L$, $E^d_H$ and energy splitting $\Delta E^d$. (c) peak area $A_L$, $A_H$ and $A_{total}$. Solid lines are guides to the eye.

To analyze the unoccupied density of states (DOS) across DW transition, we plotted the integrated area $A_L$, $A_H$ and $A_{total}$ of $a$ peak in Fig. 4(c), which closely relates to electrical transport properties for Ni $d_{x^2-y^2}$ orbital derived bands. For La327, $A_L$ shows nearly temperature-independent but a strong temperature-dependent behaviour for $A_H$. Especially, $A_H$ shows a fast decrease below 150 K, indicating a decline of unoccupied DOS from $d_{x^2-y^2}$ orbital. These are consistent with the extracted maximum peak height data [see Fig. 6(a)]. According to ref. [40], we infer that the synergetic expansion of $b$-axis and reduction of Ni-O-Ni bond angle along $b$-axis weakens in-plane hybridization between Ni $d_{x^2-y^2}$ and oxygen's $p_x/p_y$ orbitals, reducing the $d_{x^2-y^2}$ bandwidth. Although there occurs a slight expansion of $c$-axis parameter (in $Amam$ setting) to weaken the interlayer interaction below 160 K, the concurrent increase of Ni-O-Ni bond angle along $c$-axis would instead enhance bilayer interaction. This should totally increase the $d_{3z^2-r^2}$ bandwidth, as supported by FWHM data [see Fig. 6(b)]. One key finding is evidenced that an anomalous enhancement of mutual interplay between $d_{x^2-y^2}$ orbital and $d_{3z^2-r^2}$ orbital exists in La2Pr327, manifesting a stronger competitive thermal response across and away from DW transition. In further, there occurs a more significant peak energy $E_{\mathrm{peak}}$ [see Fig. 7] and area decline of $A$ peak below $T_{\mathrm{dw}}$ for La2Pr327, signalling the key role of Pr substituents in varying electronic band structure and lattice distortions. It seems that magnetic Pr substituents could intensively activate the mutual interplay between Ni's $d_{x^2-y^2}$ and $d_{3z^2-r^2}$ orbitals mediated through a hybridization of La/Pr $5d$ and O $2p$ orbitals [35]. Whether the magnetic $4f$ orbitals are also at play calls for further study to elucidate the underlying mechanism. Overall, total area $A_{\mathrm{total}}$ decreases below 150 K suggests the decline of unoccupied DOS across DW transition for both samples, which is not incompatible to the opening of energy gap at $E_{\mathrm{F}}$ and the vanishing of $d_{3z^2-r^2}$ bonding band [44].

## III. DISCUSSIONS

Bulk SC is observed in compressed $La_2PrNi_2O_{7-\delta}$ [25], in which Pr-substitution suppresses the intergrowth of La327/La4310 and La327/La214 phases that are assumed to be detrimental to bulk SC. From theoretical perspective, bilayer structure is one crucial factor for SC in pressurized La327 [7,45]. It creates an interlayer superexchange AFM interaction of $d_{3z^2-r^2}$ electrons through inner apical oxygen's $p_z$ orbital to induce local spin singlets with large pairing energy in strong-coupling limit [7], meanwhile, the increased orbital hybridization with the more itinerant $d_{x^2-y^2}$ orbital establishes phase coherence. Other strong-coupling theory considers onsite Hund's coupling of Ni with forming an interorbital spin-triplet state as well as the interlayer superexchange AFM interaction of $d_{3z^2-r^2}$ electrons in the limit $J_{\mathrm{H}} \gg J_{\parallel,\perp}$ ($J_{\mathrm{H}}$ is Hund's coupling, $J_{\parallel,\perp}$ represents intralayer and interlayer AFM spin-exchange) [42,46]. Through Hund's coupling, $J_{\perp}$ could provide strong interlayer pairing to enhance pairing strength in $d_{x^2-y^2}$ orbital [42,46-48]. In this context, one can reasonably assume that activation of $d_{3z^2-r^2}$ electrons is one beneficial factor for triggering bulk SC. For La2Pr327, our observation of thermally-promoted mutual interplay of $d_{x^2-y^2}$ orbital and $d_{3z^2-r^2}$ orbital warrant further high-pressure XANES study to reveal their role in emerged bulk SC in bilayer RP-nickelates.

## IV. CONCLUSIONS

In summary, we reported results of temperature-dependent Ni $K$-edge XANES for polycrystalline $La_3Ni_2O_{7-\delta}$ and $La_2PrNi_2O_{7-\delta}$ bilayer RP-nickelates, unraveling their orbital signatures across DW transition. Our results demonstrate a clear energy anomaly across DW transition from the Ni $4p$ and $3d$ orbitals energy splitting, signifying the occurrence of lattice distortions. Combined with the well-established SDW transition at ~150 K from the multiple experimental evidences reported, our results would support an intertwined ordered state of SDW and CDW in $La_3Ni_2O_{7-\delta}$ and $La_2PrNi_2O_{7-\delta}$. We further show that both $d_{x^2-y^2}$ and $d_{3z^2-r^2}$ orbitals are involved and mutually competitive across DW transition, below which crystal field splitting energy is enlarged due to enhanced $NiO_6$ octahedral distortion. Importantly, we reveal that Pr-substitution strengthens the mutual interplay of $d_{x^2-y^2}$ and $d_{3z^2-r^2}$ orbitals, which warrants further study to uncover its underlying mechanism and its impact on bulk SC emergence.

## ACKNOWLEDGEMENTS

W.G.Y. and N.N.L. acknowledge support from the National Nature Science Foundation of China under grants No. U2230401 and 12204022, respectively. Y.P.Q. acknowledges support from the National Natural Science Foundation of China under grant No. 52272265 and the National Key R&D Program of China under grant No. 2023YFA1607400. We acknowledge synchrotron SOLEIL for allocating beamtime at the ODE beamline (Saint-Aubin, France) under proposal No. 20231950 and the staff of ODE beamlines for assistance with data collection.

## AUTHOR CONTRIBUTIONS

M.T.L. conceived and designed this project. Y.P.Q. and W.G.Y. supervised this project. M.X.Z., C.Y.P. and Y.P.Q. synthesized and characterized the $La_3Ni_2O_7$ and $La_2PrNi_2O_7$ polycrystalline samples. M.T.L. and M.X.Z. analyzed crystal structure parameters, electrical transport, and magnetic susceptibility data. N.-D.A., and Q.Y.K. contributed to experimental set-up at ODE beamline, synchrotron SOLEIL. M.T.L., N.N.L., and J.Y.G. measured the XANES data. M.T.L. analyzed the XANES data and performed theoretical simulations. Y.M.W. supported DFT-based calculations. M.T.L. wrote the manuscript with input from all authors.

# END MATTER

## A. XANES spectra and corresponding derivatives

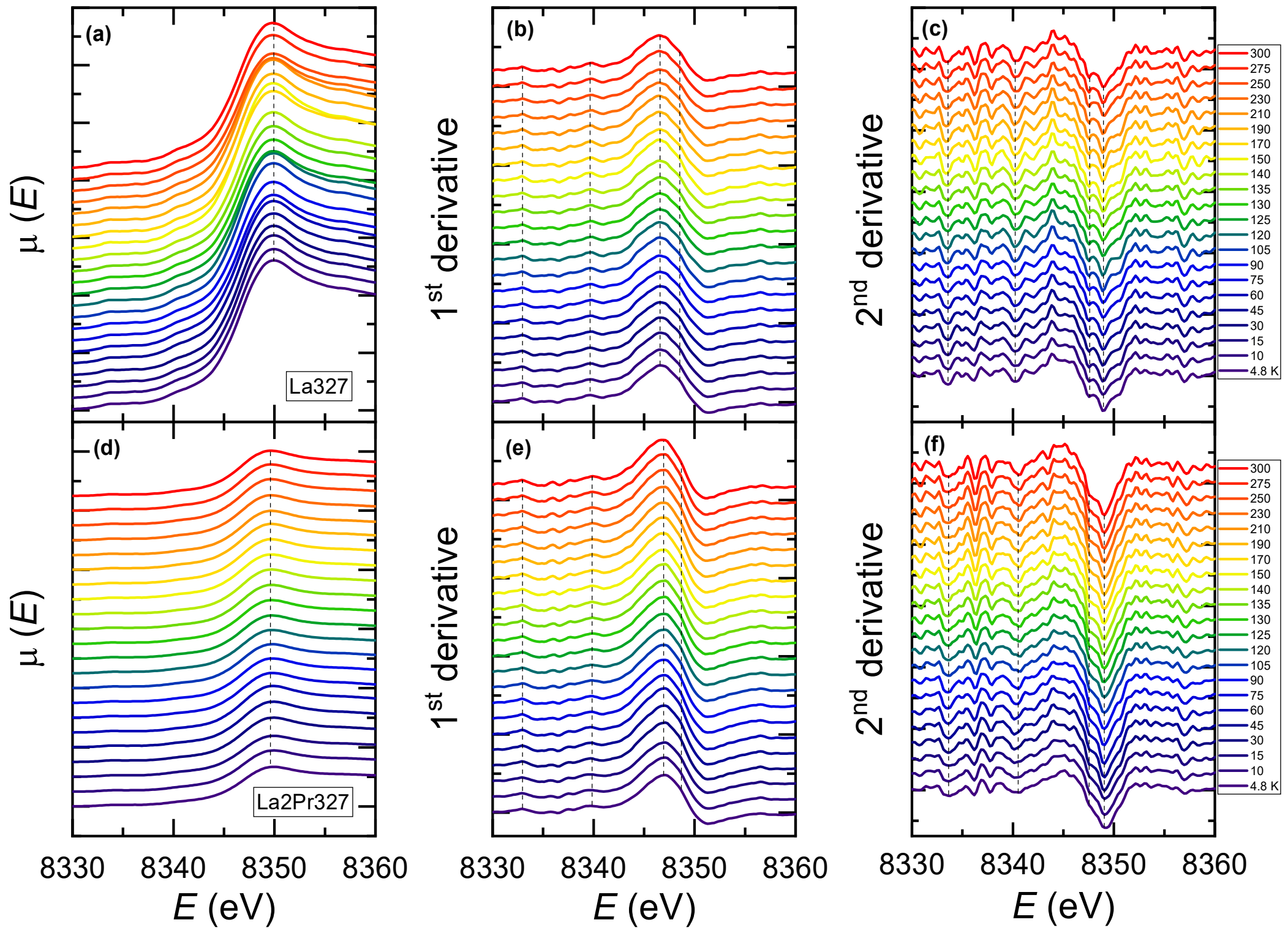

**FIG. 5. XANES, 1st derivative, 2nd derivative. (a)-(c) for La327, and (d)-(f) for La2Pr327.** Dashed lines are guides to the eye.

## B. DW-induced resistivity anomaly: role of bandwidth variation

Previous $R_{\mathrm{H}}(T)$ reveals an anomalous increase below ~115 K in La327 [11,12], which is sensitive to oxygen vacancies content. Below 115 K, dramatic increase of $R_{\mathrm{H}}$ indicates a Fermi surface reconstruction or variations of electron/hole pocket, signaling a decrease of itinerant carrier density. Optical conductivity study shows the $d_{3z^2-r^2}$ bonding band sinks below $E_{\mathrm{F}}$ and an energy gap is also evidenced when temperature lowers below $T^* \cong$115 K [44]. Since $d_{3z^2-r^2}$ bonding band is more localized and contributes to a small hole pocket [42,43], its sinking alone is hard to explain large $R_{\mathrm{H}}$ increase. According to Drude model, resistivity is given by $\rho = m^*\tau^{-1}/e^2n \propto m^*\tau^{-1}R_{\mathrm{H}}$ with $m^*$ is carrier effective mass, $\tau^{-1}$ scattering rate, and $n$ carrier density. To experimentally estimate the evolution of electrotonic correlation strength across DW transition, we extracted FWHM for $\alpha$ peak [see Fig. 6(b)], which is propositional to bandwidth. We find that $\mathrm{FWHM_L}$ and $\mathrm{FWHM_H}$ corresponding to dominant antibonding $d_{3z^2-r^2}$ and $d_{x^2-y^2}$ orbitals mutually compete to form a concave and convex-like behavior across DW transition. This implies that the electron correlation strength, thus $m^*$, follow opposite trending, namely, first decreases then increases for $d_{x^2-y^2}$ band. Taking into account the monotonous evolution of $\tau^{-1}$ across DW transition [44], we propose that the concave anomaly of resistivity is mainly ascribed to in-plane lattice distortion induced bandwidth variations [Fig. 6(c) and (d)], which reduces hole-pocket and enlarges electron-like pocket of $d_{x^2-y^2}$ band [41,49]. This could naturally explain $R_{\rm H}$ increase below $T_{\rm dw}$.

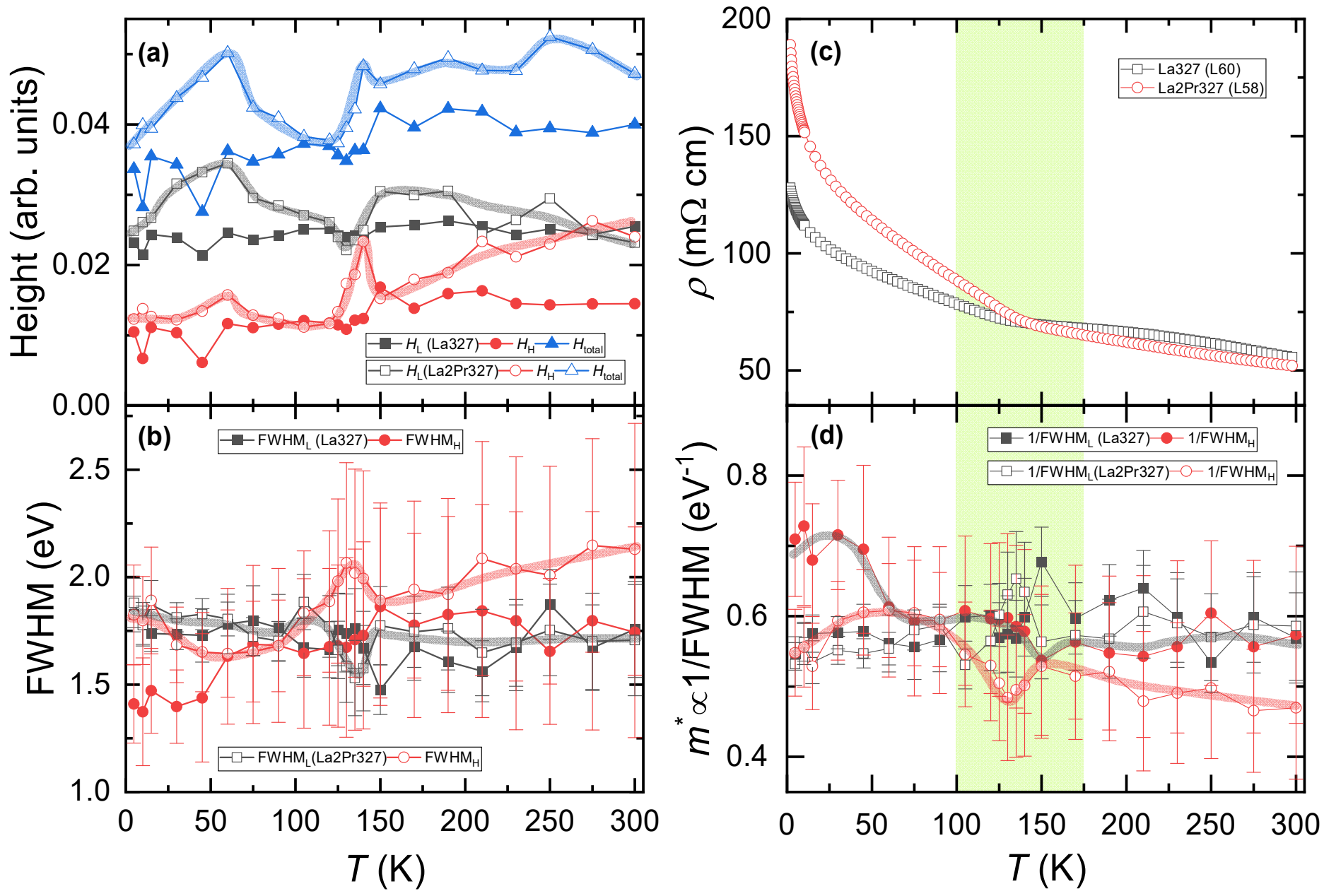

**FIG. 6. Temperature-dependent maximum height and FWHM of $a$ peak.** (a) Height. (b) FWHM. (c) $\rho(T)$. (d) 1/FWHM. Solid bold lines are guides to the eye.

### C. Orbital hybridization of La/Pr $5d$ states and O $2p$ states

Stronger and distinct electronic-lattice coupling is evidenced from the thermal cooling induced variations of $A$ peak area and energy in La2Pr327, which are indicators of unoccupied states of Ni $4p$ orbitals enhanced through La/Pr $5d$ and O $2p$ orbital hybridizations and lattice distortion affected via La/Pr ions. Below ~150 K, clear negative correlation of peak area and $E_{\rm peak}$ is evidenced for La327 and La2Pr327. A decrease of area just below DW transition signals the weakened La/Pr $5d$ and O $2p$ orbital hybridizations. According to our DFT-based calculations on La327 [35], all $5d$ orbitals contributed to La-O hybridization but only $d_{3z^2-r^2}$ and $d_{x^2-y^2}$ orbitals are relevant to the $A$ peak. The bond angles of La(1,1)-O(4)-La(1,2)/La(2,1)-O(4)-La(2,2) and La(2,3)-O(2)-Ni are mostly correlated to the strength of in-plane and out-of-plane La-O-Ni hybridizations. Based on the results of ref. [40], we have extracted the temperature-dependent atomic distances and bond angles [Fig. S5(a) in ref. [27]]. On one hand, the decrease of La(2,3)-O(2)-Ni angle would reduce the orbital hybridization of $d_{3z^2-r^2}$ and $p_z$ to result in DOS decline thus decrease of $A$ peak area [Fig. S5(b)]. On one other hand, the concomitant increase and decrease of La(1,1)-O(4)-La(1,2) and La(2,1)-O(4)-La(2,2) would instead neutralize their hybridization contribution to retain the in-plane DOS unchanged. However, the anomalous evolution of its edge energy should be controlled by multiple factors, in which the atomic distances among La(1)/La(2) sites, Ni and O(2)/O(4) sites need to be considered [Fig. S5(c)].

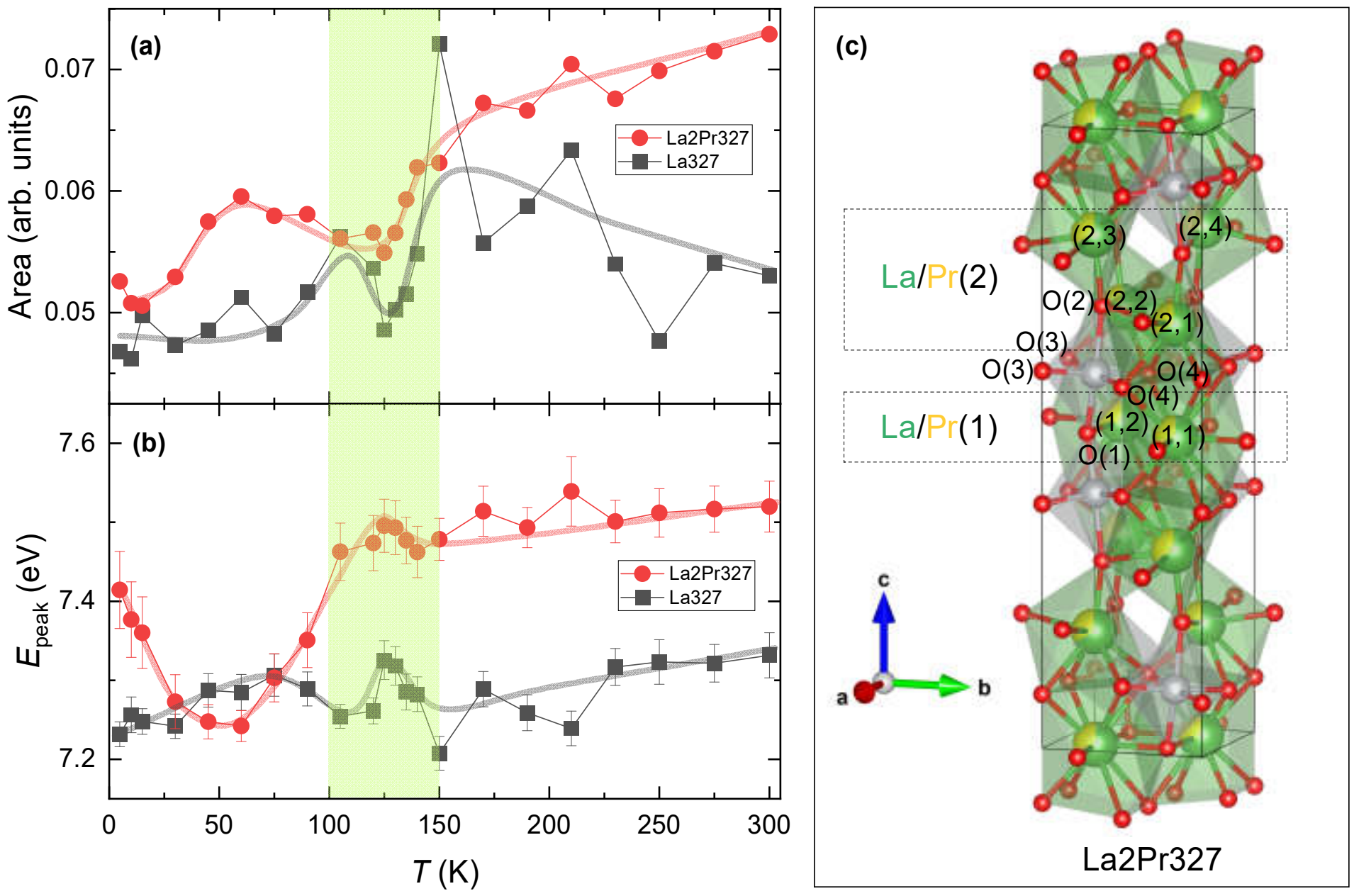

**FIG. 7. Temperature-dependent integrated area and $E_{\mathrm{peak}}$ of $A$ peak.** (a) Area. (b) $E_{\mathrm{peak}}$. (c) Crystal structure with $LaO_8$/$PrO_8$ polyhedron and $NiO_6$ octahedron. Solid bold lines are guides to the eye.

## Supplementary materials

# I. Methods

## Sample synthesis, characterizations, magnetic susceptibility and electrical transport measurements

Polycrystalline samples of $La_3Ni_2O_{7-\delta}$ and $La_2PrNi_2O_{7-\delta}$ were prepared by sol-gel method [1]. Stoichiometric amount of $La(NO_3)_3\cdot 6H_2O$ (AR, 99%, Aladdin), $Pr_6O_{11}$ oxide first dissolved in nitric acid, and $Ni(NO_3)_2\cdot 4H_2O$ (99.9%, Aladdin) were dissolved into water, which was heated before forming a transparent, light green solution. Then, citric acid (AR, ≥99.5%, Aladdin) was added. After evaporating the water, the obtained green gel was heated to 800 °C for 6 h to produce black powders, which were pressed into dense pellets for sintering at 1150 °C for 48 h. Powder X-ray diffraction (XRD) patterns were collected on a Bruker D2 phaser with Cu-$K_\alpha$ radiation (step <0.02° for 2theta scan) at room temperature. Rietveld refinement was performed by GSAS II package [2]. Chemical composition is semi-quantitively determined by energy-dispersive X-ray spectrometry (EDS). Electrical transport and magnetic properties were measured on the Physical Property Measurement System (Dynacool, Quantum Design) and SQUID vibrating sample magnetometer (MPMS3, Quantum Design).

**Table S1** Lattice parameters ($a$, $b$, $c$), volume ($V$), Ni-O bond distances ($d_1$, $d_2$, $d_3$, $d_4$, $\langle d \rangle$), Ni-O-Ni bond angles, and $NiO_6$ distortion parameter (Δ) of $La_3Ni_2O_{7-\delta}$ and $La_2PrNi_2O_{7-\delta}$ polycrystalline samples.

| Samples | $a$ (Å) | $b$ (Å) | $c$ (Å) | $V$ (Å$^3$) | $\alpha$ (°) | $\beta$ (°) | Ref. |
|---|---|---|---|---|---|---|---|
| | $d_1$ (Å) | $d_2$ (Å) | $d_3$ (Å) | $d_4$ (Å) | $\langle d \rangle$ (Å) | Δ | |
| La327 | 5.39840 | 5.44944 | 20.50620 | 603.25666 | 178.82300 | 167.46060 | This work |
| | 1.96105 | 2.24074 | 1.91456 | 1.93725 | 2.01340 | 0.00352 | |
| La327 | 5.39283 | 5.44856 | 20.51849 | 602.89803 | 178.86130 | 167.64490 | [3] |
| | 1.97468 | 2.22707 | 1.90648 | 1.94101 | 2.01231 | 0.00330 | |
| La2Pr327 | 5.39064 | 5.46208 | 20.37070 | 599.79707 | 178.81260 | 167.34890 | This work |
| | 1.94830 | 2.22609 | 1.91523 | 1.93813 | 2.00694 | 0.00322 | |
| La2Pr327 | 5.38206 | 5.46208 | 20.43430 | 600.71207 | 178.20000 | 164.20000 | [4] |
| | 1.98300 | 2.23000 | 1.92300 | 1.93000 | 2.01650 | 0.00324 | |

## Theoretical XAFNES simulations

Theoretical XANES simulations were performed by finite difference method near edge structure (FDMNES) code [5,6], which is mainly a fully relativistic DFT-LSDA (local spin density approximation) code allowing $ab\ initio$ simulations [7]. In our calculations of Ni K-edge XANES, full-potential finite difference method (FDM) was used to solve the electronic structure. Self-consistent DFT approach was adopted with considering SOC for the heavy La and Pr atoms, which naturally make our calculations relativistic. Electronic configurations of the absorber Ni are set to be $3d^8 4s^2$ for initial state, and $3d^8 4s^2 4p^1$ for excited state. Lattice parameters are used from experimental reported for $Amam$ symmetry of La327 [3] and La2Pr327 [4]. Cluster size of 7.0 Å is set to calculate the XANES ranging from -50−100 eV. Two sets of XANES with quadrupole and without quadrupole channels were calculated to evaluate the absolute contribution of quadrupole.

**XAFNES measurements**

**TABLE S2** Energy scale of $a$ peak and $B'/B$ peaks extracted from the theoretical simulated and experimental (4.8/300 K) XANES. All energy values are relative to 8333 eV.

| Sample | Symmetry | Peak | $E_{\mathbf{edge}}$ (eV) | $E_{\mathrm{L}}$ (eV) | $E_{\mathrm{H}}$ (eV) | $\Delta E$ (eV) | $\Delta E_{\mathbf{exp.}}$ (eV) |
|---|---|---|---|---|---|---|---|
| La327 | $Amam$ | $a$ | / | 0.86(1) | 1.90(7) | 1.04(4) | 1.48/1.49 |
| | | $B'/B$ | 12.64(1) | 13.02(4) | 15.16(2) | 2.13(3) | 1.45/1.40 |
| La2Pr327 | $Amam$ | $a$ | / | 0.76(2) | 1.89(4) | 1.14(3) | 1.38/1.16 |
| | | $B'/B$ | 12.56(1) | 13.13(3) | 15.22(1) | 2.10(2) | 1.47/1.40 |

X-ray absorption near edge spectroscopy (XANES) measurements were conducted at ODE beamline [8], synchrotron SOLEIL. ODE is dedicated to energy-dispersive HP extended XAFS and X-ray magnetic circular dichroism (XMCD) experiments between 3.5 keV to 25 keV [8]. Si(111) was also used in our HP measurements with the same energy resolution $\Delta E/E$~$1.4\times10^{-4}$ eV (~1.17 eV at the Ni K-edge) but great energy stability due to the absence of any mechanical movement. Typical energy step of 0.22 eV per pixel was obtained after calibrating the XAS data collected at ODE. Typical beam spot size is about 25 μm ×35 μm in FWHM with tail of 70 μm (at 7 keV). Optical cryostat was used to conduct the low temperature (5−300 K) XAFS measurements with an open cycling liquid helium, which is transmitted by a liquid helium transfer line between He Dewar and cryostat. Transmission mode was adopted for XAFS data acquisition at RT and low temperature. Fine powders were freshly ground from dense pellets and compressed into a pre-indented stainless sample chamber. Nickel foil standard was measured for energy calibration.

## II. Curie-Weiss fitting result

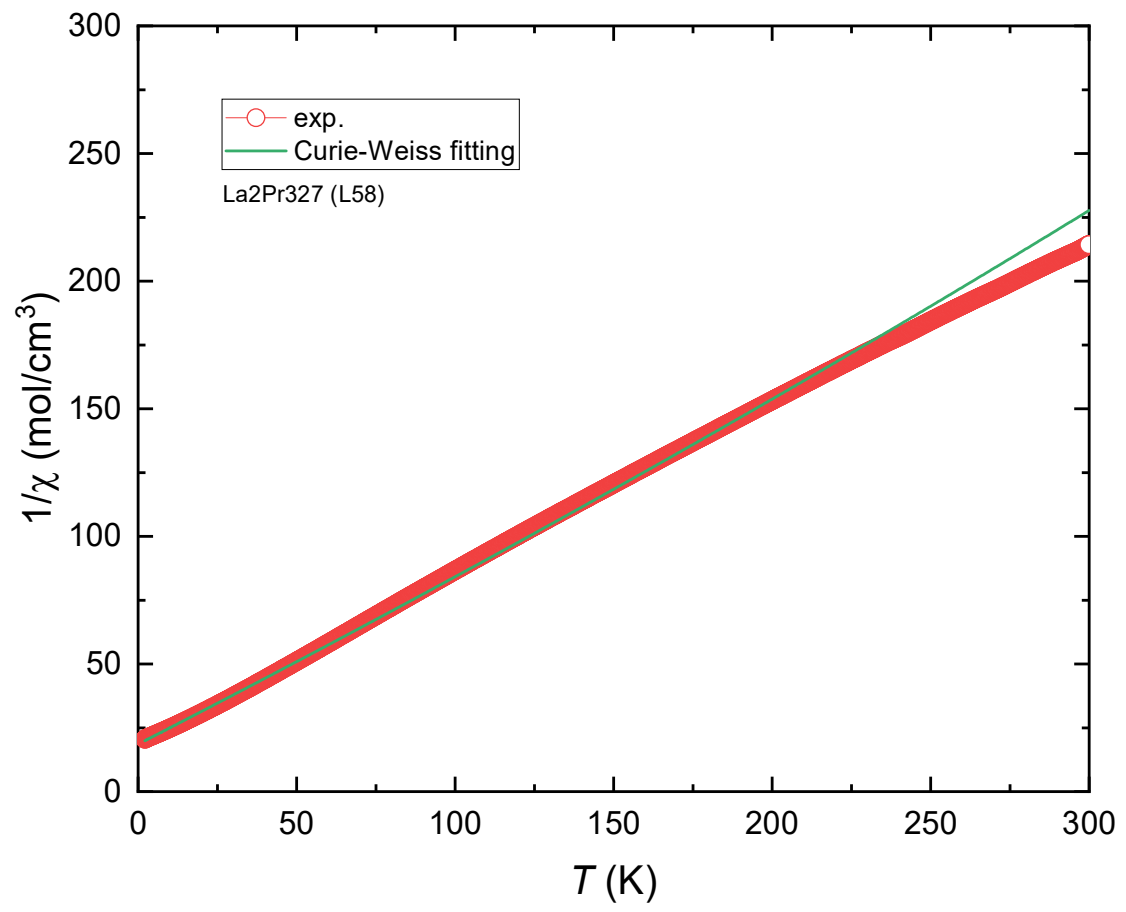

**FIG. S1.** Curie-Weiss fitting profile of $\chi(T)$ for La2Pr327 sample.

## III. $a$ peak and $A$ peak at various temperature

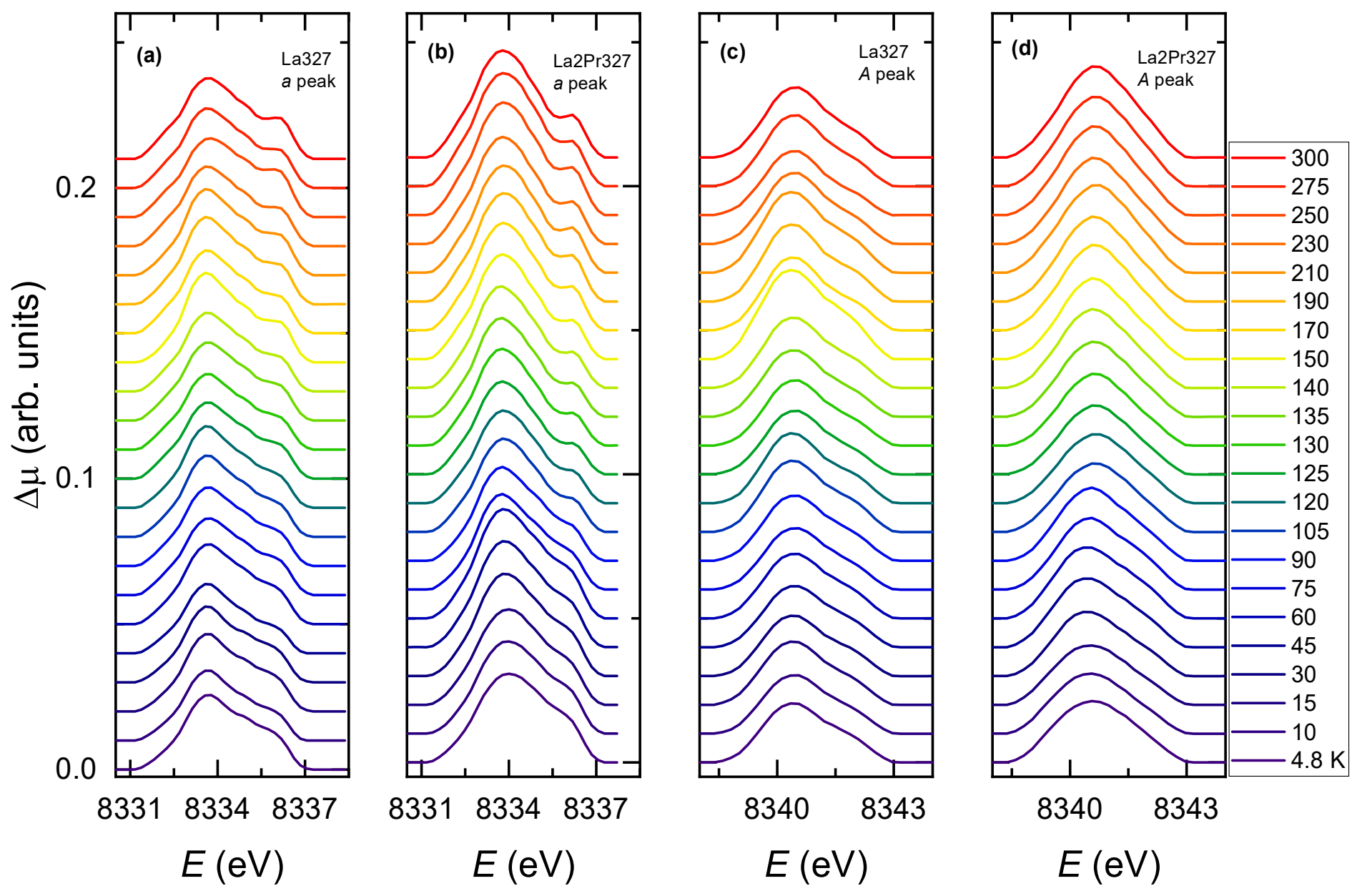

**FIG. S2.** Temperature-dependent $a$ peak and $A$ peak for La327 and La2Pr327 samples.

## IV. Lattice distortions and Ni-O-Ni bonding parameters at low temperature

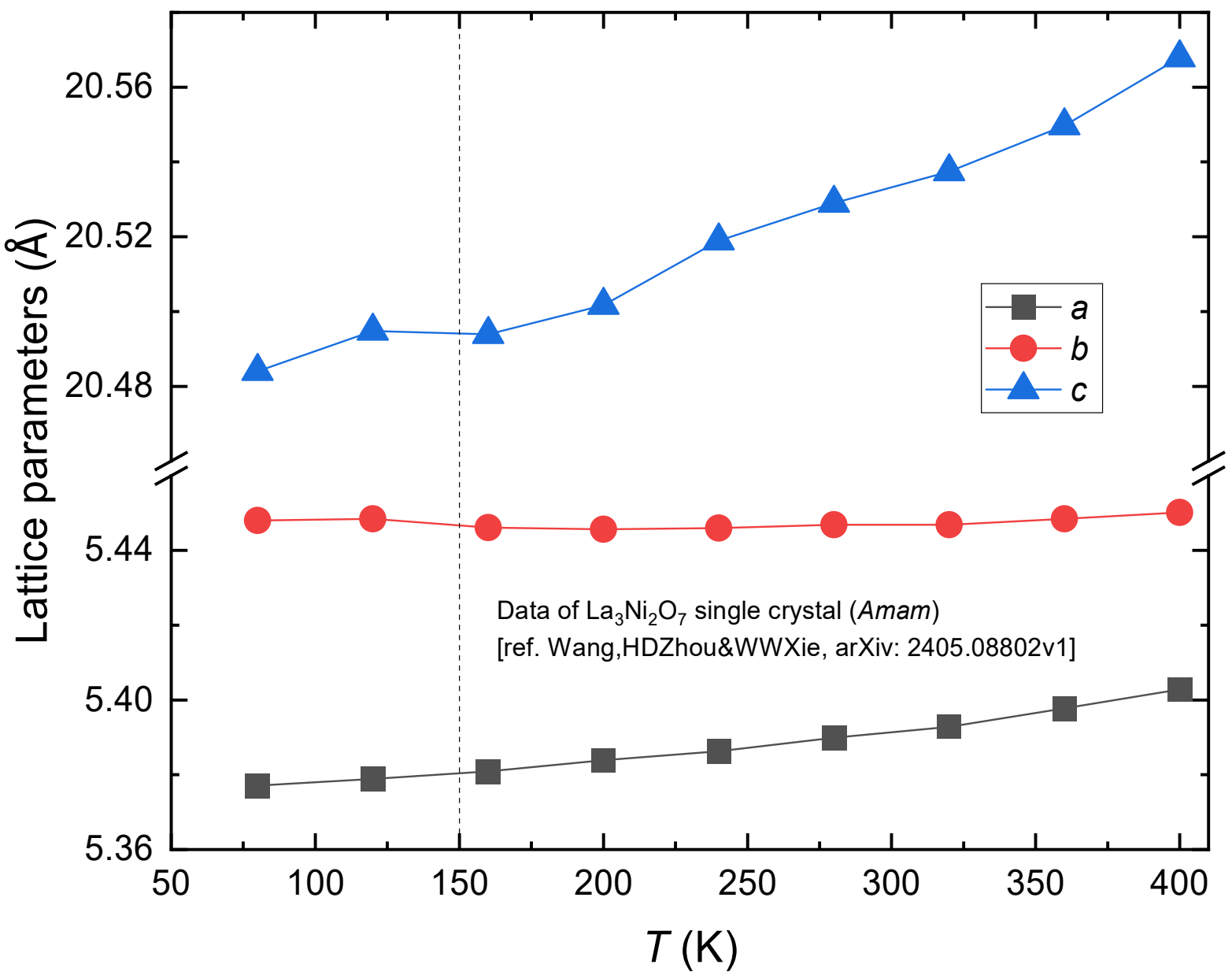

**FIG. S3.** Lattice parameters obtained from La327 single crystal [9].

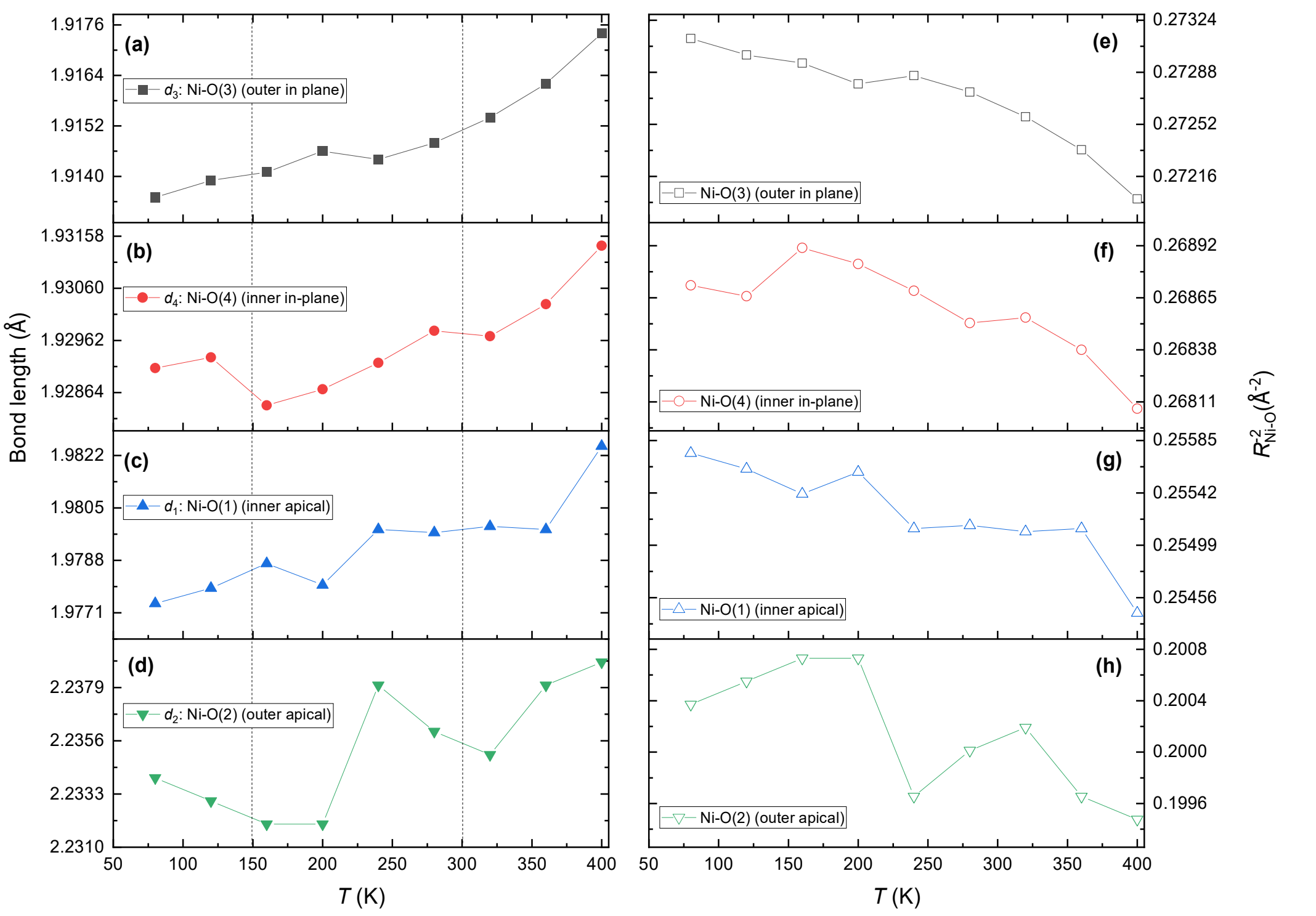

**FIG. S4. Ni-O bond length and its inversed squared $R_{\mathrm{Ni-O}}^{-2}$ obtained from La327 single crystal [9].** (a) Ni-O bond distances as a function of temperature. There are four inequivalent Ni-O bond distances in $Amam$ phase. (b) Corresponding $R_{\mathrm{Ni-O}}^{-2}$ values.

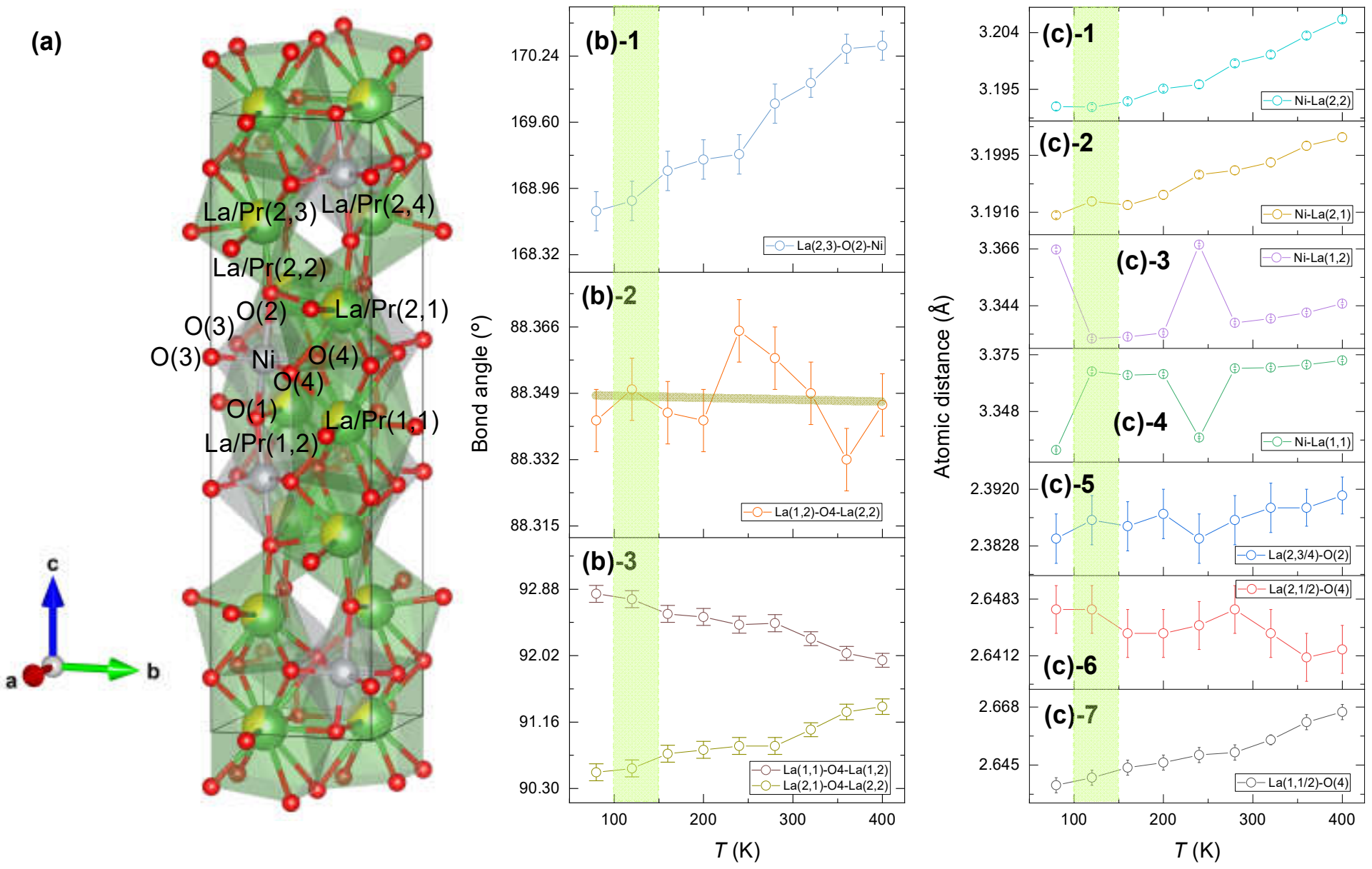

**FIG. S5. La-Ni-O bond length and angles obtained from La327 single crystal [9].** (a) Schematic crystal structure of La2Pr327. (b) La-O-Ni and La-O-La bond angles as a function of temperature. (c). Ni-La and La-O atomic distances as a function of temperature.